\documentclass[%
 reprint,
 amsmath,amssymb,
 aps,
]{revtex4-1}

\usepackage{graphicx}
\usepackage{dcolumn}
\usepackage{bm}

\begin{document}

\preprint{This line only printed with preprint option}

\title{Global quantum discord in infinite quantum spin chains}

\author{Zhao-Yu Sun,$^{1\dagger}$ Yan-E Liao,$^1$ Bin Guo,$^2$ Hai-Lin Huang,$^1$ Duo Zhang,$^1$  Jian Xu,$^1$ Bi-Fu Zhan,$^1$
Yu-Yin Wu,$^1$ Hong-Guang Cheng,$^1$
Guo-Zhi Wen,$^1$, Chao Fang,$^1$
Cheng-Bo Duan,$^3$ Bo Wang$^4$}

\affiliation{$^1$School of Electrical and Electronic Engineering, Wuhan Polytechnic University, Wuhan 430000, China.\\
$^2$School of Science, Wuhan University of Technology, Wuhan 430070, China.\\
$^3$College of Science, Chang'an University, Xi'an 710064, China.\\
$^4$ENN Group Co., Ltd., Langfang, Hebei 065001, China.\\
$^\dagger$ corresponding author: sunzhaoyu2000@gmail.com}




\begin{abstract}
In this paper, we study global quantum discord (GQD) in infinite-size spin chains.
For this purpose,  in the framework of matrix product states (MPSs),  we propose an effective
procedure to calculate GQD (denoted as $G_n$) for consecutive $n$-site subchains
in infinite chains. 
For a spin-$\frac{1}{2}$ three-body interaction model, whose ground state can be
exactly expressed as MPSs,
We use the procedure to study $G_{n}$  with $n$ up to $24$.
Then for a spin-$\frac{1}{2}$ XXZ chain, we firstly use infinite time-evolving block decimation (iTEBD)
algorithm to obtain the approximate wavefunction in the from of MPSs, and then figure out $G_n$ with $n$ up to $18$.
In both models, $G_{n}$ shows an interesting linear growth as the increase of $n$, that is, $G_{n}\sim k\cdot n+b$.
Moreover, in non-critical regions the slope $k$ of $G_n$ converges very fast, while in critical regions it converges relatively slow,
and the behaviors are explained in a clear physical picture with the short-range and long-range correlations.
Based on these results, we propose to use $\frac{G_n}{n}$  to describe the global correlations in infinite chains.
$\frac{G_n}{n}$ has twofold physical meanings. Firstly, it can be regarded as ``global discord per site'',
very similar to ``energy per site'' or ``magnetization per site'' in quantum magnetic systems. Secondly, 
$\frac{G_n}{n}$ (when $n$ is large enough) describes the quantum correlation between a single site and an ($n$-1)-site block.
Then we successfully apply our theory to an
exactly soluble infinite-size spin XY chain which is beyond the matrix product formula, and the Hamiltonian can reduce to the
transverse-field Ising model and the XX model.
The relation between GQD and quantum phase transitions in these models is discussed.
\end{abstract}
\maketitle

\section{introduction}

Quantum phase transitions (QPTs) are driven by zero-temperature quantum
fluctuations.\cite{BOOK_QPT} For a many-body system described by a Hamiltonian $\hat{H}(g)$,
where $g$ can be an external perturbation(e.g., the magnetic field) or an internal interaction constant,
the ground state of the system may show qualitative change when the
external perturbation $g$ crosses some $g_{c}$. Then one says a
QPT occurs, with $g_{c}$ the QPT point. Recently, concepts from quantum
information theory, such as quantum entanglement, have been used to
characterize QPTs.\cite{QE_QPT,Bell_inequalitiesQPTs_XXZ_model,Bell_QPT_models,QE1,multi_entanglement,Camp_GQD_chain} The entanglement entropy, which describes the
quantum entanglement between a subchain with its environment, has
been studied in many one-dimensional quantum spin chains, and an interesting
scaling behavior has been observed.\cite{QE1} For instance, in a spin XY model,
as the increase of the subchain, entanglement entropy shows a logarithmic growth
in the vicinity of the QPT point meanwhile achieves a finite saturation value in non-QPT
regions.\cite{XY3} The scaling behavior is found to be related to the long-range
(short-range) correlations of the system at (off) the QPT point. These
studies have deepened our understanding of QPTs.

With the help of Bell-type inequalities, quantum entanglement has
been generalized into multi-partite settings,\cite{MBI1,MBI2,MBI3} where a concept---the
hierarchy of correlations--- emerges naturally in many-body
systems. It has been found that the XY model shows high (low) hierarchy
of multipartite correlation at (off) the second-order QPT point.\cite{MBImy} Furthermore,
we have studied the multipartite correlation in the XXZ model, and
observed low hierarchy of multipartite correlation at the infinite-order
QPT point of the chain.\cite{tobepub} The result is unexpected, since a system usually
shows maximum mount of quantum correlation at the QPTs. A possible
reason is that, the QPT in the XXZ model is driven by other type of
quantum correlation, rather than the entanglement.

An alternative  of entanglement is quantum discord (QD),
since QD can be used to capture all the quantumness of correlation
in a quantum two-qubit state, with entanglement included.\cite{QD1,QD2,QD3} QD has
been investigated extensively in many quantum models, such as one-dimensional
quantum spin chains and matrix product systems.\cite{QD_chain1,QD_chain2,my2D_QDchain,myQDChain} Recently, the concept
of QD has been generalized into multi-site settings, which is called
global quantum discord (GQD).\cite{GQD1,GQD2,GQD3,GQD_mono1,GQD_review}

The behavior of GQD in finite-size spin chains have been studied in
several papers, and the relation between GQD and QPTs has been discussed.\cite{Camp_GQD_chain,GQD_chain}
For instance, GQD show a maximum in the vicinity of the QPT point of
the transverse-field Ising model.\cite{Camp_GQD_chain} We need to mention that, in all
the models studied (including the transverse field Ising model, the
cluster-Ising model, the spin XX model and the Ashkin-Teller model),
(i) the length of the finite-size chain is very short, i.e., $N\le12$, and
(ii) GQD does not show any clue of convergence as the increase of $N$.
Thus, an effective method to characterize GQD in infinite-size spin chain
is still unknown. Since QPT is a macroscopic phenomena occurring in infinite-size
systems rather than finite-size systems, the study of GQD in infinite chains would be valuable.

In this paper, we will study GQD in infinite-size spin chains. In
the framework of matrix product states (MPSs),\cite{MPS} we propose an effective
procedure to calculate GQD (denoted as $G_n$) for consecutive $n$-site subchains
in one-dimensional infinite-size spin chains. Since the ground states
of one-dimensional translational-invariant chains can always be approximately expressed
as MPSs through the infinite time-evolving block decimation (iTEBD)
algorithm, our procedure can be widely adopted in quantum spin chains.\cite{iTEBD1,iTEBD2}
We use the method to study $G_{n}$ for an infinite three-body interaction model with
$n$ up to $24$
and an infinite XXZ model with $n$ up to $18$.
In both models, for any given physical parameter, as the increase
of $n$, $G_{n}$ shows an interesting linear growth, that is, $G_{n}\sim k\cdot n+b$.
Moreover, in non-critical regions $k$ converges very fast, while in critical regions it converges relatively slow.
The behaviors are explained in a clear physical picture with the short-range and long-range correlations.
Then we successfully apply our theory to an
exactly soluble infinite-size spin XY chain, which can reduce to the
transverse-field Ising model and the XX model.

This paper is organized as follows. In Sec, II, we briefly review
the concept of GQD.
A three-body interaction model with MPSs as the ground state is studied in Sec. III.
A spin XXZ model is studied with the help of iTEBD algorithm in Sec. IV.
An exactly soluble spin XY model is discussed in Sec. V.
A summary is given in Sec. VI.

\section{concepts and formula}

\subsection{Global quantum discord}

First of all, let's define the von Neumann entropy and the relative entropy.\cite{GQD1} For a general state $\rho$, the von Neumann entropy is defined as
\[
S(\rho)=-Tr[\rho\log_{2}\rho].
\]
For two general states $\rho$ and $\sigma$ which lie on
the same Hilbert space, one can further define the relative entropy as
\[
S(\rho||\sigma)=Tr[\rho\log_{2}\rho]-Tr[\rho\log_{2}\sigma].
\]

Next, we briefly review the concept of two-site symmetric quantum discord
$G_{2}$.\cite{GQD1} For a two-site system (consisting of site $A_{1}$ and site $A_{2}$)
described by a density matrix $\rho_{A_{1}A_{2}}$, we will use $\rho_{A_{j}}$
to denote the reduced density matrix for site $A_j$. Then the symmetric
discord of the system can be defined in terms of the relative entropy
as

\begin{equation}
\begin{array}{ccc}
G_{2}(\rho_{A_{1}A_{2}}) & = & \min\{S(\rho_{A_{1}A_{2}}||\Phi_{A_{1}A_{2}}(\rho_{A_{1}A_{2}}))\\
 &  & -\sum_{j=1}^{2}S(\rho_{A_{j}}||\Phi_{A_{j}}(\rho_{A_{j}}))\}
\end{array}.\label{eq:G2}
\end{equation}
In the above expression, $\Phi_{A_{j}}(\rho_{A_{j}})$ denotes a locally
projective measurement performing on site $A_j$, i.e., $\Phi_{A_{j}}(\rho_{A_{j}})=\sum_{l}\Pi_{l_{j}}\rho_{A_{j}}\Pi_{l_{j}}$,
with \{$\Pi_{l_{j}}$\} the set of projectors.  $\Phi_{A_{1}A_{2}}(\rho_{A_{1}A_{2}})$
denotes a local multi-site measurement on the entire system, i.e.,
\[
\Phi_{A_{1}A_{2}}(\rho_{A_{1}A_{2}})=\sum_{l_{1}l_{2}}(\Pi_{l_{1}}\otimes\Pi_{l_{2}})\rho_{A_{1}A_{2}}(\Pi_{l_{1}}\otimes\Pi_{l_{2}}).
\]
The minimization in Eq. (\ref{eq:G2}) is
according to all the projectors.
If $\rho_{A_{1}A_{2}}$ contains no quantum correlation, one can prove
that $G_{2}$ is equal to zero. If $\rho_{A_{1}A_{2}}$ contains any
form of quantum correlation(such as entanglement), $G_{2}$ will be
a positive number. Thus $G_{2}$ captures all the content of non-classical
correlations in $\rho_{A_{1}A_{2}}$.

Global discord can be regarded as a direct generalization of the two-site
quantum discord $G_{2}$.\cite{GQD1} We consider an
$n$-site density matrix $\rho_{A_{1}...A_{n}}$.
The global discord for $\rho_{A_{1}...A_{n}}$ can be defined as

\begin{equation}
\begin{array}{ccc}
G_{n}(\rho_{A_{1}...A_{n}}) & = & \min\{S(\rho_{A_{1}...A_{2}}||\Phi_{A_{1}...A_{2}}(\rho_{A_{1}...A_{2}}))\\
 &  & -\sum_{j=1}^{n}S(\rho_{A_{j}}||\Phi_{A_{j}}(\rho_{A_{j}}))\}
\end{array},\label{eq:org_GQD}
\end{equation}
where the minimization is according to all the local projectors, and
\[
\Phi_{A_{1}...A_{2}}(\rho_{A_{1}...A_{2}})=\sum_{l_{1}...l_{n}}(\otimes_{j=1}^{n}\Pi_{l_{j}})\rho_{A_{1}...A_{n}}(\otimes_{j=1}^{n}\Pi_{l_{j}}).
\]
$G_{n}$ can be used to characterize all the quantumness of correlations
in the multi-site state $\rho_{A_{1}...A_{n}}$.

In very few special situations(for instance,
the Werner-GHZ state), analytical expressions of $G_{n}$ have been
obtained.\cite{GQD_analytic}
In more general situations,  $G_{n}$ should be numerically figured out.
Recently, Campbell has reformulated
the formula in Eq. (\ref{eq:org_GQD}) as\cite{Camp_GQD_chain}

\begin{equation}
\begin{array}{ccc}
G_{n}(\rho_{A_{1}...A_{n}}) & = & \min\{\sum_{j=1}^{n}\sum_{l=1}^{2}\tilde{\rho}_{j}^{ll}\log_{2}\tilde{\rho}_{j}^{ll}\\
 &  & -\sum_{\mathbf{l}=1}^{2^{n}}\tilde{\rho}^{\mathbf{ll}}\log_{2}\tilde{\rho}^{\mathbf{ll}}\}\\
 &  & +\sum_{j=1}^{n}S(\rho_{A_{j}})-S(\rho_{A_{1}...A_{n}})
\end{array}.\label{eq:Camp_GQD}
\end{equation}
Here
\begin{equation}
\tilde{\rho}_{j}^{ll}=\langle l|R_{j}^{\dagger}\rho_{A_{j}}R_{j}|l\rangle,\label{eq:rho_l}
\end{equation}
with $j=1,...,n$ labeling the sites, $l=1,2$ denoting the spin down
and spin up states, and $R_{j}=\left(\begin{array}{cc}
\cos\frac{\theta_{j}}{2} & \sin\frac{\theta_{j}}{2}e^{-i\phi_{j}}\\
\sin\frac{\theta_{j}}{2}e^{i\phi_{j}} & -\cos\frac{\theta_{j}}{2}
\end{array}\right)$ is the local rotation for site $j$. Moreover,
\begin{equation}
\tilde{\rho}^{\mathbf{ll}}=\langle\mathbf{l}|\mathbf{R}^{\dagger}\rho_{A_{1}...A_{n}}\mathbf{R}|\mathbf{l}\rangle,\label{eq:rho_L}
\end{equation}
with $\mathbf{l}=1,...,2^{n}$ denoting the standard basis of $n$-site
Hilbert space, and
\begin{equation}
\mathbf{R}=\otimes_{j=1}^{n}R_{j}\label{eq:big_R}
\end{equation}
is a  multi-site  local rotation. $S(\cdot)$ is
the von Neumann entropy. Despite its complex form, Eq. (\ref{eq:Camp_GQD})
is much more efficient in numerical calculations than the original formula in
Eq. (\ref{eq:org_GQD}). In fact, several finite-size systems with
up to 11 sites have been investigated.\cite{Camp_GQD_chain} For larger systems, nevertheless, the calculation still becomes
difficult.

\subsection{matrix product formula}

In this section, we will show how to calculate global discord efficiently
for continuous $n$-site subchains
in infinite matrix product states (MPSs). There are many reasons why we should
study MPSs.\cite{MPS} Firstly, there are some quantum models, whose ground state
can be exactly expressed as MPSs.\cite{MPS} In addition, MPSs captures the basic
correlation properties of many gapped systems, that is, short-range
correlation. The results obtained in MPSs may reflect the behavior
of global discord in other models which are difficult (or impossible) to solve.
Furthermore, for a general one-dimensional quantum system, the ground
state can always be express approximately as MPSs with
the help of infinite time-evolving block decimation(iTEBD) algorithm.\cite{iTEBD1}
As we will show, we are able to figure out the global discord
for continuous $n$-site subchains in infinite
systems with $n$ up to 24.

An $N$-site spin-$\frac{1}{2}$ MPS with a periodic boundary condition
is expressed as\cite{MPS}

\begin{equation}
|\psi\rangle=\sum_{j_{1}j_{2}j_{3}...j_{N}}\textrm{Tr}(M_{j_{1}}M_{j_{2}}M_{j_{3}}\cdots M_{j_{N}})|j_{1}j_{2}j_{3}...j_{N}\rangle,\label{eq:MPS}
\end{equation}
where $j_i=1,2$ denote the spin-down state and the spin-up state
of the $i$-th spin, and $\{M_{j}\}$ are $D\times D$ matrices, which are site-independent.
We only consider infinite systems, that is, $N\rightarrow\infty$.
For a few models, the ground states can be exactly expressed
as MPSs with $D=2$.
For general quantum models,  the MPSs can also describe
the ground states with very high accuracy
when $D$ is large enough.\cite{MPS_faith}

For a given MPS, the reduced density matrix $\rho_{A_{1}...A_{n}}$
of any continuous $n$-site subchain can be calculated conveniently.
Firstly, we re-write Eq. (\ref{eq:MPS}) as

\begin{equation}
|\psi\rangle=\sum_{\mathbf{j_{n}}=1,2^{n}}\sum_{j_{n+1}...j_{N}}\textrm{Tr}(\mathbf{S}_{\mathbf{j_{n}}}M_{j_{n+1}}\cdots M_{j_{N}})|\mathbf{j_{n}}j_{n+1}...j_{N}\rangle\label{eq:MPS_1}
\end{equation}
with $|\mathbf{j_{n}}\rangle=|j_{1}...j_{n}\rangle$ according to
the $2^{n}$ basis of the $n$-site Hilbert space, and
\[
\mathbf{S}_{\mathbf{j_{n}}}=M_{j_{1}}...M_{j_{n}}.
\]
For an infinite-size MPS, the density matrix $\rho_{A_{1}...A_{n}}$ of a continuous $n$-site block can be figured out according to
\begin{equation}
\langle\mathbf{k}|\rho_{A_{1}...A_{n}}|\mathbf{m}\rangle=\langle\lambda|\mathbf{S}_{\mathbf{k}}^{*}\otimes\mathbf{S}_{m}|\lambda\rangle,\label{eq:rho_general}
\end{equation}
where $\mathbf{k,m}=1,...,2^{n}$ denote the standard basis of $n$-site
Hilbert space, $^*$ denotes the complex conjugate, and
 $\langle\lambda\vert$
and $\vert\lambda\rangle$
are the left  and right eigenvector
 of the transfer matrix $T=\sum_{j}M_{j}^{*}\otimes M_{j}$,
according to
its largest eigenvalue  $\lambda$.\cite{MPS}

We will show how to figure out $\tilde{\rho}_{j}^{ll}$
and $\tilde{\rho}^{\mathbf{l}\mathbf{l}}$ efficiently for a given
multi-site  local rotation operator
$\mathbf{R}=\otimes_{j=1}^{n}R_{j}$.
We will suppose $R_{j}$ is site-independent, that is, $R_{j}=R$.
In the case $R_j$ is site-dependent, very slight modifications are needed in the following formula.

(a) Calculation of $\tilde{\rho}_{j}^{ll}$. Firstly, by inserting
identity operators $I=\sum_{k=1}^{2}|k\rangle\langle k|$ and $I=\sum_{m=1}^{2}|m\rangle\langle m|$
into Eq. (\ref{eq:rho_l}), we obtain

\[
\tilde{\rho}_{j}^{ll}=\sum_{k,m=1,2}\langle l|R_{j}^{\dagger}|k\rangle\langle k|\rho_{A_j}|m\rangle\langle m|R_{j}|l\rangle.
\]
The expression for $\langle k|\rho_{A_j}|m\rangle$ can be figured out according to the
general formula in Eq. (\ref{eq:rho_general}), with $n=1$. Let's define
\[
d_{l}=\sum_{n=1,2}M_{n}\langle n|R|l\rangle,
\]
 and
\[
e_{l}=d_{l}^{*}\otimes d_{l}.
\]
After straightforward calculations, one will find that

\begin{equation}
\tilde{\rho}_{j}^{ll}=\langle\lambda|e_{l}|\lambda\rangle.\label{eq:rhoj_ll}
\end{equation}

(b) Calculation of $\tilde{\rho}^{\mathbf{l}\mathbf{l}}$. We insert
identity operators $\mathbf{I}=\sum_{\mathbf{k}=1}^{2^{n}}|\mathbf{k}\rangle\langle\mathbf{k}|$
and $\mathbf{I}=\sum_{\mathbf{m}=1}^{2^{n}}|\mathbf{m}\rangle\langle\mathbf{m}|$
into Eq. (\ref{eq:rho_L}), and obtain

\[
\tilde{\rho}^{\mathbf{l}\mathbf{l}}=\sum_{\mathbf{k},\mathbf{m}=1}^{2^{n}}\langle\mathbf{l}|\mathbf{R}^{\dagger}|\mathbf{k}\rangle\langle\mathbf{k}|\rho_{A_{1}...A_{n}}|\mathbf{m}\rangle\langle\mathbf{m}|\mathbf{R}|\mathbf{l}\rangle.
\]
The elements of $\rho_{A_{1}...A_{n}}$ has already been figured out in Eq. (\ref{eq:rho_general}).
In addition, keep in mind that $\mathbf{R}$ is the direct product
of local rotation operators (see Eq. (\ref{eq:big_R})), thus the
explicit expressions for $\langle\mathbf{l}|\mathbf{R}^{\dagger}|\mathbf{k}\rangle$
and $\langle\mathbf{m}|\mathbf{R}|\mathbf{l}\rangle$ are also available.
Finally, we find that

\begin{equation}
\tilde{\rho}^{\mathbf{l}\mathbf{l}}=\langle\lambda|e_{l_{1}}\cdot e_{l_{2}}\cdot...\cdot e_{l_{n}}|\lambda\rangle.\label{eq:rho_ll}
\end{equation}

The advantage of our formula in Eqs. (\ref{eq:rhoj_ll}) and (\ref{eq:rho_ll}) is that, we do not need to store the
$2^{n}\times2^{n}$ density matrix $\rho_{A_{1}...A_{n}}$. Instead, we just need
to store the two $D^{2}\times D^{2}$ matrices $e_{l}$.

Then we can carry out the optimization in Eq. (\ref{eq:Camp_GQD})  effectively by scanning
the rotation operator $R$,
which is the most time-consuming part in the calculation of GQD.

(c) The last one problem is the calculation of the von Neumann entropy
$S(\rho_{A_{1}...A_{n}})$, where the full eigenvalue spectrum of $\rho_{A_{1}...A_{n}}$
is needed. When $n$ is very large, the storage and exact diagonalization
of $\rho_{A_{1}...A_{n}}$ becomes impossible, thus some approximation
algorithm is needed. Here we propose to use the idea of density matrix
renormalization group (DMRG) algorithm, with some slight modification.\cite{DMRG}
Our goal is to find some approximate wavefunction of Eq. (\ref{eq:MPS}),
that is,
\begin{equation}
|\psi\rangle\approx\sum_{\tilde{\mathbf{j_{x}}}=1,\tau}\sum_{j_{x+1}...j_{N}}\textrm{Tr}(\widetilde{\mathbf{S}}_{\mathbf{\tilde{j_{x}}}}M_{j_{x+1}}\cdots M_{j_{N}})|\tilde{\mathbf{j_{x}}}j_{x+1}...j_{N}\rangle,\label{eq:MPS_reduced}
\end{equation}
where $\tilde{\mathbf{j_{x}}}=1,...,\tau$ denote the basis of the
Hilbert space of the first $x$ sites in the chain (the so-called
``system block'' in DMRG language). In practice, $\tau\ll2^{x}$
so that the density matrix of the ``system block'', a $\tau\times\tau$
matrix(see Eqs. (\ref{eq:MPS_1}) and (\ref{eq:rho_general})), can be easily diagonalized. For such a purpose, we first figure
out the reduced density matrix $\rho_{A_{1}...A_{x}}$ of the ``system
block'' according to Eq. (\ref{eq:rho_general}), and construct a transform
matrix $U_{\mathbf{j_{x}},\tilde{\mathbf{j_{x}}}}$. The columns of
$U_{\mathbf{j_{x}},\tilde{\mathbf{j_{x}}}}$ are just the eigenvectors
of $\rho_{A_{1}...A_{n}}$, according to the first $\tau$ largest eigenvalues.
Then the MPS in Eq. (\ref{eq:MPS_1}) is reduced into Eq. (\ref{eq:MPS_reduced}),
with

\[
\widetilde{\mathbf{S}}_{\mathbf{\tilde{j_{x}}}}=\mathbf{S}_{\mathbf{j_{x}}}\cdot U_{\mathbf{j_{x}},\tilde{\mathbf{j_{x}}}}.
\]

In practice, we start the renormalization with a ``system block'' containing
$4$ sites, i.e, $x=4$. We reduce the dimension of the Hilbert space
to $\tau$ , then enlarge the system block to contain $x+1$ sites.
Repeat the renormalization steps until the target $n$ is finally
reached.

\section{a simple mps model}

We consider a three-body interaction infinite-size model described by the following
Hamiltonian\cite{MPS}
\[
\hat{H}=\sum_{i}J_{3}\sigma^{z}_{i}\sigma^{x}_{i+1}\sigma^{z}_{i+2}+J_{z}\sigma^{z}_{i}\sigma^{z}_{i+1}-B\sigma_{i}^{x},
\]
where $\sigma_{i}^{\alpha}$ with $\alpha=x,y,z$ denote the Pauli
matrices on site $i$. $J_3$, $J_z$ and $B$ are interaction constants.
It has
been found that when $J_{3}=(g-1)^{2}$, $J_{z}=2(g^{2}-1)$, and
$B=(1+g)^{2}$, the ground state can be exactly described by MPSs
with $M_{1}=\left(\begin{array}{cc}
0 & 0\\
1 & 1
\end{array}\right)$ and $M_{2}=\left(\begin{array}{cc}
1 & g\\
0 & 0
\end{array}\right)$.
When $g=-1$, the system just contains three-body interactions and its ground state is the so-called cluster state.\cite{Cluster}
On the other hand, when $g=1$, one can see that the system is in a fully polarized state along the $x$ direction. Thus there should be a QPT between
 $g=-1$ and  $g=1$. Further studies show that the magnetization in the $x$ direction keeps nonzero for $g>0$
and vanishes for $g<0$, and the QPT occurs at $g_c=0$.\cite{MPS}

\begin{figure}
\includegraphics[scale=1]{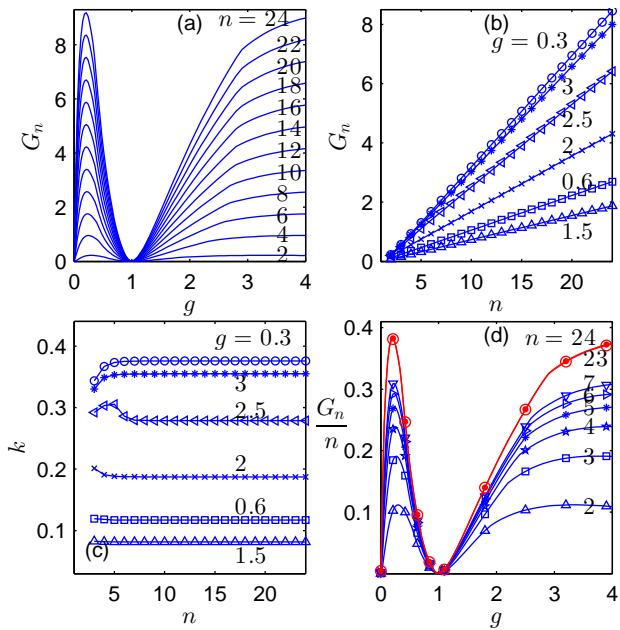}\caption{\label{fig:MPS} (Color online) Global discord $G_n$ of continuous $n$-site subchain in the infinite MPS model.
(a) The dependence of $G_n$ on $g$ with $n$ up to 24.
(b) The dependence of $G_n$ on $n$ for several fixed $g$.
(c) The fast convergence of the slope of $G_n(n)$ with the increase of $n$, where $k=G_n-G_{n-1}$ for fixed $g$.
(d) The dependence of the discord per site $\frac{G_n}{n}$ on $g$ for several $n$.
}
\end{figure}

The results of global discord for continuous $n$-site subchains  with $n$ up to 24 are shown in Fig. \ref{fig:MPS}.
In Fig. \ref{fig:MPS}(a), firstly, we find that the first-order derivative of $G_n$ would be divergent at $g_c=0$ when $n$ is large enough, which should result from the dramatic change of the ground state. Thus  $G_n$ signals the QPT of this model.
Secondly, $G_n$ is zero at $g=1$. Since the system is in a fully polarized state, it is obvious that there should not be any quantum correlation in the system thus $G_n=0$. At the QPT point $g_c=0$, though the interactions in the system are very complex, our result $G_n=0$ discloses that the system contains no quantum correlation and should be expressed as a product
(or factorized) state, too. It shows that $G_n$ is valuable in investigating factorization of quantum states in complex models.\cite{fac}

As the increase of $n$, $G_n$ increases steadily and there is no clues for convergence. Thus it would be a valuable question that, how should we describe the global discord in an infinite chain?
As illustrated in Fig. \ref{fig:MPS}(b), $G_n$ shows an approximately linear dependence upon the size of the subchain for any $g$, i.e.,
\[
G_n\sim k\cdot n + b.
\]
To see the linearity more clearly, we present the dependence of the slope $k$ upon $n$ in Fig. \ref{fig:MPS}(c), where $k(n)=\frac{G_n-G_{n-1}}{n-(n-1)}=G_n-G_{n-1}$. As the increase of $n$, $k$ converges
fast. It indicates that the curves in Fig. \ref{fig:MPS}(b) indeed can be regarded as lines when $n$ is large enough. In the limit $n\rightarrow\infty$, we will have
$\frac{G_n}{n}=k$, thus $\frac{G_n}{n}$ would also converge.  Up to now $\frac{G_n}{n}$ has a more clear physical meaning than $k$, i.e., it denotes the ``global discord per site'', very similar to ``energy per site'' and ``magnetization per site'' in quantum magnetic chains. In Fig. \ref{fig:MPS}(d) we show the behavior of discord per site as a function of $g$ for several $n$. One can see that $\frac{G_n}{n}$ converges very fast.
In fact, it becomes difficult to distinguish the curves for $n=23$ and $n=24$ from each other.

It is interesting that whether the above results are specific to this MPS model, or is general in other one-dimensional quantum spin chains.
Thus we will reconsider the issue in an infinite spin-$\frac{1}{2}$ XXZ model.

\section{xxz model}

An one-dimensional infinite spin-$\frac{1}{2}$ XXZ model is described by
the following Hamiltonian\cite{Bell_inequalitiesQPTs_XXZ_model,BOOK_QPT}

\[
\hat{H}=\sum_{i}\sigma_{i}^{x}\sigma_{i+1}^{x}+\sigma_{i}^{y}\sigma_{i+1}^{y}+\Delta\sigma_{i}^{z}\sigma_{i+1}^{z},
\]
where $\sigma_{i}^{\alpha}$ denote the Pauli
matrices, and $\Delta$ is the anisotropic parameter.
In the limit $\Delta\rightarrow+\infty$ and $\Delta\rightarrow-\infty$,
the model should be in an anti-ferromagnetic phase and a ferromagnetic phase, respectively.
In the intermediate region $-1\le\Delta\le1$ the system is
in a gapless phase, and there is a first-order QPT at $\Delta_c=-1$
and an infinite-order QPT at $\Delta_c=+1$.
The Hamiltonian has two symmetries, i.e., a U(1) rotation about the $z$ axis
and a $Z_2$ symmetry about the $x$ or $y$ axis.
The $Z_2$ symmetry can be broken when $|\Delta|\ge1$.

The analytical formula of the ground-state energy per site ($e_0(\Delta)$) of the system can be found in Ref..
Suppose the $Z_2$ symmetry is unbroken,
then for two nearest-neighboring sites $A_1$ and $A_2$, all the elements of the reduced density matrix  $\rho_{A_{1}A_{2}}$ can be obtained according to the derivative of the energy, $\frac{\partial{e_0(\Delta)}}{\partial{\Delta}}$.
As a result, exact result for two sites are available in some papers.

For large $n$, it is difficult to identify the reduced density matrices $\rho_{A_{1}...A_{n}}$.
Thus we use the infinite time-evolving block decimation (iTEBD) algorithm to express
approximately the ground state of the infinite XXZ
model by matrix product states. Details about the iTEBD algorithm can be found
in Refs. [21, 22].  Finally,
the global discord can be figured out conveniently by the formula for MPSs proposed in Sec. II(B).
It need mention that when $|\Delta|\ge1$,  the  $Z_2$ symmetry is broken spontaneously in the iTEBD algorithm, since a random initial state is used in the evolution.

We set $D=16$ in our iTEBD calculations.
To check the accuracy of the algorithm, we have compared the approximate
ground-state energy from iTEBD algorithm with the exact ground-state energy from Ref. .
The relative error turns out to be smaller than $1.6\times10^{-4}$ for any $\Delta$.

\begin{figure}
\includegraphics{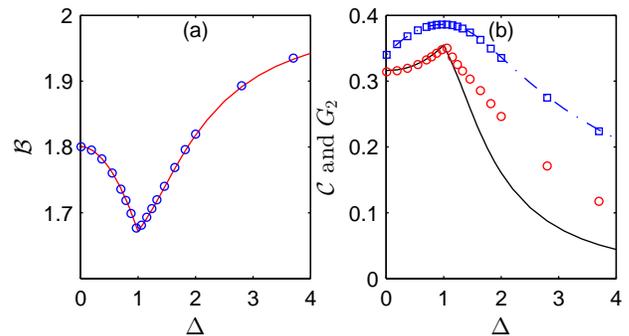}\caption{\label{fig:XXZ_n2} (Color online) Results of quantum correlations for two nearest-neighboring sites in the infinite XXZ model.
(a) Numerical results (circles) and analytical results (line) for Bell inequality.
(b) Numerical results (squares) and analytical results (dash line) for entanglement concurrence,
and numerical results (circles) and analytical results (solid line) for discord.}
\end{figure}

Furthermore, we compare our numerical results (with $Z_2$ symmetry broken when $|\Delta|\ge1$) of some two-site quantum correlations with
previously reported exact results(with $Z_2$ symmetry unbroken).
We consider the two-site Bell inequality(A measure of the so-called quantum nonlocality. The definition and exact results for the XXZ model can be found in Ref. \cite{Bell_inequalitiesQPTs_XXZ_model}),
the two-site concurrence(A measure of quantum entanglement. The definition and exact results for the XXZ model can be found in Ref. \cite{CC_SSB}), and the global discord.
The violation measure $\mathcal{B}$ of the Bell inequality is shown in Fig.  \ref{fig:XXZ_n2}(a),
and the concurrence  $\mathcal{C}$  and global discord $G_2$ are shown in Fig.  \ref{fig:XXZ_n2}(b).
One can see that for Bell inequality and concurrence, our numerical results are in good consistence with the analytical results.
For the discord, a clear discrepancy can be found when $\Delta>1$ (where the $Z_2$ symmetry is broken spontaneously). It suggests that the nonlocality
and entanglement are unaffected by spontaneous symmetry breaking,
meanwhile the discord is sensitive to spontaneous symmetry breaking.
The behavior is consistent with previous studies
about the effect of symmetry-breaking on two-site entanglement\cite{CC_SSB} and two-site discord\cite{discord_SSB}.
Finally, the validity of our program is confirmed.

\begin{figure}
\includegraphics[scale=1]{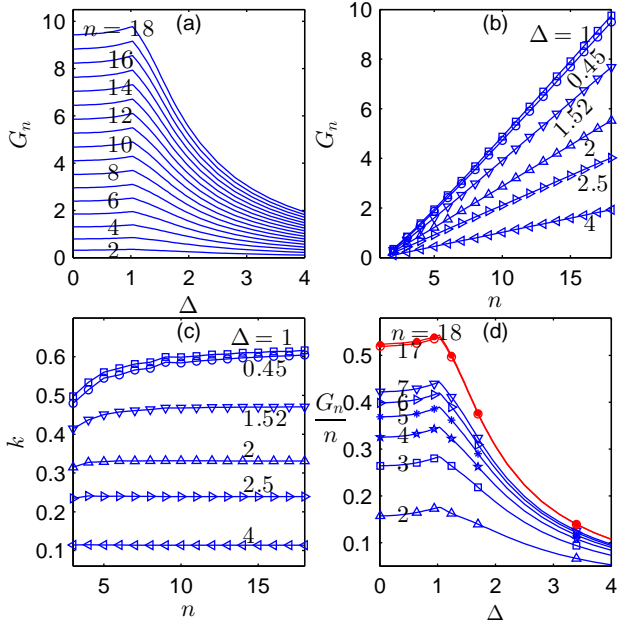}\caption{\label{fig:XXZ}  (Color online) Global discord $G_n$ of continuous $n$-site subchain in the infinite XXZ model.
(a) The dependence of $G_n$ on $\Delta$ with $n$ up to 18.
(b) The dependence of $G_n$ on $n$ for several fixed $\Delta$.
(c) The fast convergence of the slope of $G_n(n)$ with the increase of $n$, where $k=G_n-G_{n-1}$ for fixed $\Delta$.
(d) The fast convergence of discord per site $\frac{G_n}{n}$.}
\end{figure}

Our results of global discord with $n=2,3,...,18$ are shown in Fig. \ref{fig:XXZ}(a).
Firstly, the global discord $G_n$ shows a size-independent peak in the vicinity of the QPT point $\Delta=1$,
which means that the system present large amount of quantum correlation at the QPT point.

\begin{figure}
\includegraphics{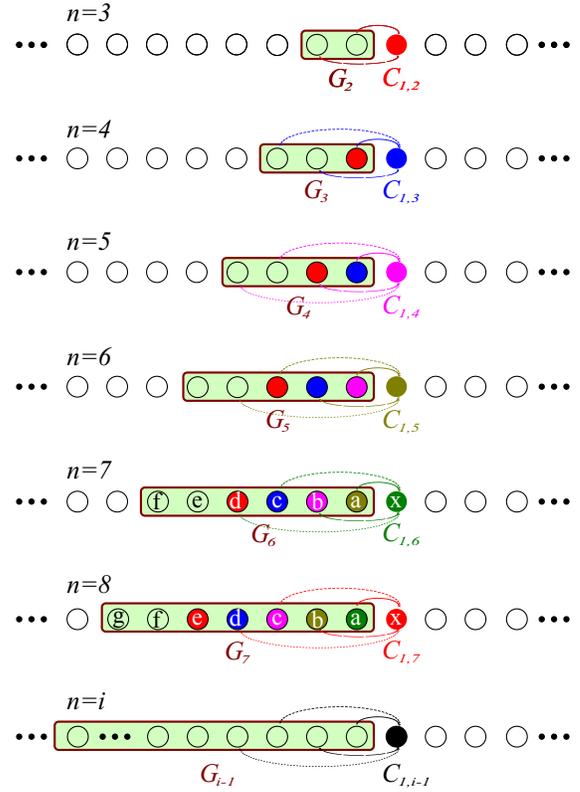}\caption{\label{fig:explain} (Color online)
A continuous $n$-site block consists of two subsystems, that is, an ($n$-1)-site block (according to the left $n-1$ sites surrounded by the rectangle)
and a single site (according to the single colored site outside the rectangle).
Under this splitting, the global correlation $G_n$ for continuous $n$ sites contains two parts, that is, the global correlation $G_{n-1}$ in the ($n$-1)-site block and the
correlation $C_{1,n-1}$ between the ($n$-1)-site block and the single site.}
\end{figure}

As the increase of $n$, $G_n$ shows no clue for convergence. In order to investigate the scaling behavior,
we show the size dependence of $G_n$ for several $\Delta$ in Fig. \ref{fig:XXZ}(b). An approximately linear behavior
is observed, just as in the above MPS model in Sec. III.
In Fig. \ref{fig:XXZ}(c) we illustrate the size dependence of the slope of $G_n$, where $k(n)=G_n-G_{n-1}$.
When $\Delta>1$, as the increase of $n$, $k$ converges quite fast. On the other side,
when $\Delta\le1$, the convergence is very slow. Finally, the discord per site is shown in Fig. \ref{fig:XXZ}(d).
One can see that $\frac{G_n}{n}$ converges fast and the curves for $n=17$ and $n=18$ overlap each other.

Since similar results have been observed in the MPS model and the XXZ model,
the above behavior may be general for one-dimensional infinite spin chains.
Thus we need to offer a physical explanation of our results.
We will concentrate on three issues. (a) Why do $\frac{G_n}{n}$ and $k$ converge when $n$ is large enough ?
(b) Why do $\frac{G_n}{n}$ and $k$ show different speeds of convergence
 in the gapped regions (for $\Delta>1$ in the XXZ model) and in the gapless regions (for $\Delta\le1$ in the XXZ model) ?
(c) Is there any further physical meaning of $\frac{G_n}{n}$ and $k$ in the context of quantum correlation ?

We start our discussion with $n=3$. Please see Fig. \ref{fig:explain}.
We divide the 3-site subchain into two parts, that is, the left 2 sites surrounded by the rectangle and the third site outside the rectangle.
Then the global correlation in the 3-site subchain  (described by $G_3$) contains two parts,
i.e.,
the correlation in the 2-site block (described by $G_2$)
and the correlation between the 2-site block and the single site(denoted as $C_{1,2}$).
It need mention that according to the monogamy inequality of quantum correlations, it may not hold that $G_3=G_2+C_{1,2}$. However, the difference between $G_3$ and $G_2$ can be used to qualitatively describe the correlation $C_{1,2}$.
Similarly, for general $n$, as illustrated  in Fig. \ref{fig:explain}, $G_n$ also contains two parts, denoted as $G_{n-1}$ and $C_{1,n-1}$.

Consequently, the difference between $G_n$ and $G_{n-1}$
 (or just the slope $k$ of $G_n$, since $k$ is equal to $G_n-G_{n-1}$)
 results from the correlation $C_{1,n-1}$ between the ($n$-1)-site block and the single site, i.e., $k=G_n-G_{n-1}\sim C_{1,n-1}$.
In order to understand the scaling behavior of the slope $k(n)$ in Figs. 1(c) and 3(c),
we need to analyze the dependence of
$C_{1,n-1}$ upon $n$.
$C_{1,n-1}$ consists of multi two-site correlations. Take $n=7$ for instance(please see Fig. \ref{fig:explain}). $C_{1,6}$ contains (i) the two-site correlation between site $x$ and its nearest neighbor $a$, denoted as $c_{x,a}$,
(ii) the two-site correlation between $x$ and its next-nearest neighbor $b$, denoted as $c_{x,b}$,
(iii) the correlation between $x$ and its next-next-nearest neighbor $c$, denoted as $c_{x,c}$, and etc.
In many situations two-site correlations are short-range, and the correlations are mainly distributed between nearer neighbors, i.e., $c_{x,a}$ $>$ $c_{x,b}$ $>$ $c_{x,c}$ $>$ $c_{x,d}$ $>$ $...$ $>$ $c_{x,f}$ $\thickapprox0$.
As $n$ increases from $7$ to $8$, the contribution of the two-site correlation $c_{x,g}$ should be taken into account.
However, since $c_{x,g}$ is negligible when compared with other two-site correlations, $C_{1,7}$ should be equal to $C_{1,6}$ approximately.
In other words, $C_{1,n}$ will converge when $n$ is large enough.
One can see that the convergence of $C_{1,n}$ is understandable with the help of short-range correlations. Consequently, the convergence of both $G_n-G_{n-1}$ and $k(n)$ becomes reasonable.

In the above discussion, the physical meaning of $k(n)$ also becomes clear, i.e., $k(n)$ can be used to describe the correlation between a single site and the ($n$-1)-site block. Furthermore, the convergent value of $k(n)$, in other words, $\lim_{n\rightarrow\infty}k(n)$, describes the correlation between a single site and an infinite chain.

When the correlations are long-range, the two-site correlation such as $c_{x,g}$ may not be negligible for finite $n$,
thus the convergence of $C_{1,n}$ would be relatively slow.
As a result, $k(n)$ and $\frac{G_n}{n}$ would also converge slowly.
In the XXZ model, the correlation is long-range in the gapless phase ($\Delta \le 1$) and short-range in the gapped phase ($\Delta>1$).
As a result, as shown in Fig. \ref{fig:XXZ}(c) and (d), $k(n)$ and $\frac{G_n}{n}$ indeed converge slower for $\Delta\le1$ than for $\Delta>1$.
One can see that the speed of convergence of $k(n)$ and $\frac{G_n}{n}$ is physically determined by long-range or short-range correlations in low-dimensional quantum spin systems.

\section{exact results of spin xy chain: beyond matrix product states}

In order to further confirm the conclusions in the previous sections, which are drawn from approximate numerical results,
we will study the global discord in an exactly soluble one-dimensional infinite quantum spin XY chain. No numerical approximation is involved in this section,
and the study is beyond the framework of matrix product states.

The infinite quantum spin XY chain can be described by the following Hamiltonian

\[
\hat{H}=-\frac{1}{2}\sum_{i}[\frac{1+\gamma}{2}\sigma_{i}^{x}\sigma_{i+1}^{x}+\frac{1-\gamma}{2}\sigma_{i}^{y}\sigma_{i+1}^{y}+\lambda\sigma_{i}^{z}],
\]
where $\gamma$ describes the anisotropy in the $x$-$y$
plane, and $\lambda$ is the magnetic field in the $z$ direction.
The system reduces to
the transverse-field Ising model and the XX model for $\gamma=1$ and $\gamma=0$, respectively.
In addition, at the critical point $\lambda_{c}=1$, a second-order QPT will occur when $\gamma\in(0,1]$
and an infinite-order QPT will occur when $\gamma=0$.

The infinite spin XY chain is exactly soluble
by introducing Majorana operators
and using Wick's theorem from quantum-field theory.
We were able to figure out the reduced density matrix $\rho_{A_{1}...A_{n}}$ for a subchain consisting of $n$ consecutive
spins with $n\le7$. More details of calculations can be found in Refs. \cite{XY3,MBImy}.
After obtaining $\rho_{A_{1}...A_{n}}$, we use formula (\ref{eq:Camp_GQD}) to calculate the discord.

\begin{figure}
\includegraphics[scale=1]{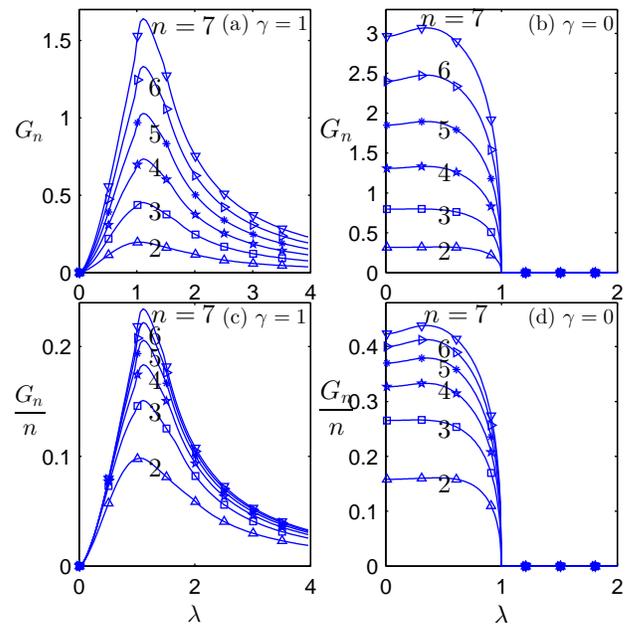}\caption{\label{fig:XY} (Color online)
Global discord for $n$-site subchains in the infinite Ising model(a) and the XX model(b).
Global discord per site for $n$-site subchains in the infinite Ising model(c) and the XX model(d).
}
\end{figure}

In Fig. \ref{fig:XY} (a) and (b), we show the results of global discord for $\gamma=1$ (the Ising model) and $\gamma=0$
 (the XX model), respectively.
For $\gamma=1$, one can see that $G_n$ will always show a peak in the vicinity of the second-order QPT point $\lambda_c=1$.
For $\gamma=0$, $G_n$ is zero for $\lambda>1$ and its first-order derivative is divergent at the infinite-order QPT point $\lambda_c=1$.
Thus global discord of subchains in the infinite chain can be very convenient to  characterize the QPTs in the XY chains.
We would like to mention that the global discord of finite-size XX chains with up to 11 sites has been studied in Ref. \cite{Camp_GQD_chain}.
One can see that $G_n$ in infinite chains, even just with $n=2$, shows more clear and sharp signal of the infinite-order QPT than in finite-size chains with up to 11 sites.

\begin{figure}
\includegraphics[scale=1]{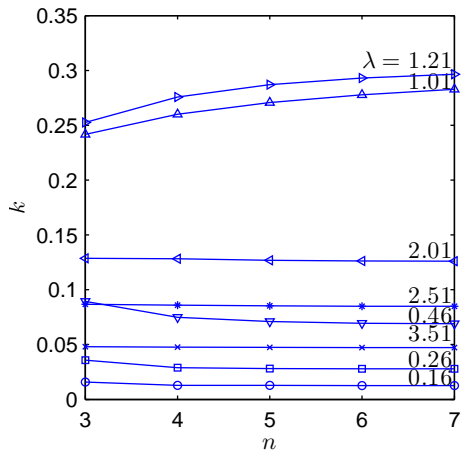}\caption{\label{fig:Ising_k} (Color online)
Convergence of $k=G_n-G_{n-1}$ in the transverse-field Ising model.
}
\end{figure}

It is obvious that as the increase of $n$, the global discord $G_n$ never converges in Fig. \ref{fig:XY}(a) and (b).
Thus we show $\frac{G_n}{n}$ for $\gamma=1$ and $\gamma=0$ in Fig. \ref{fig:XY}(c) and (d), respectively.
One can find a clear clue of convergence. Especially, in the gapped regions $\lambda\ne1$ of the transverse-field Ising model,
$\frac{G_n}{n}$ converges very fast(see Fig. \ref{fig:XY}(c)). On the other hand, in the vicinity of the gapless region $\lambda\approx1$ of the transverse-field Ising model (Fig. \ref{fig:XY}(c))
and $\lambda<1$ of the XX model (Fig. \ref{fig:XY}(d)), where the system is long-range correlated, $\frac{G_n}{n}$ converges slowly.
In Fig. \ref{fig:Ising_k} we show the convergence of $k(n)=G_n-G_{n-1}$ for the transverse-field Ising model.
One can see that $k$ converges more slowly in the vicinity of the QPT point $\lambda_c=1$ than in non-critical regions.
Thus, our exact results in these XY chains are consistent with the conclusions from numerical results in previous sections.

\section{summary and discussions}

In this paper, we have studied global discord $G_n$ for $n$-continuous spins in one-dimensional infinite spin chains.
We have proposed useful formula to calculate global discord in the framework of matrix product states (MPSs).
As the first example, we have studied $G_n$ with $n$ up to 24 in a three-body interaction model, whose ground state can be
exactly expressed as  MPSs. Then in a spin XXZ model, with the help of iTEBD algorithm we have successfully figured out  $G_n$ with $n$ up to 18.
Finally, we have considered an exactly soluble spin XY chain beyond the framework of MPSs, where its Hamiltonian can be reduced into the transverse-field Ising model and the XX model.

In all the above models, we have found some similar behaviors. Firstly,
global discord never converges as the increase of $n$, thus the scaling behavior is completely different
from the entanglement entropy which has been introduced in Sec. I.
Secondly, $G_n$ shows an approximately linear behavior as the increase of $n$, i.e., $G_n \approx k\cdot n + b$,
and its slope $k$ would converge when $n$ is large enough.
Since $\frac{G_n}{n}=k$ in the limit $n\rightarrow\infty$, we have used $\frac{G_n}{n}$ to characterize the global discord in infinite chains.
$\frac{G_n}{n}$ (and $k$) has twofold physical meanings. It can be regarded as ``global discord per site'',
very similar to ``energy per site'' in quantum spin systems. Moreover, $k(n)$
describes the quantum correlation between a single site and an ($n$-1)-site block.
The convergent value of $k$ (and $\frac{G_n}{n}$)
describes the quantum correlation between a single site and an infinite chain.
Furthermore, we find $\frac{G_n}{n}$ (and $k$) converges faster in the gapped (non-critical) regions than in the gapless (critical) regions.
The speed of convergence is determined by the short-range correlations and long-range correlations in the two situations.

The relation between quantum correlation and QPTs is an interesting topic recently.
As illustrated in the figures of this paper, the global discord shows clear signal for the QPTs.
We would like to pay our attention to the infinite-order QPT in the XXZ model(Fig. \ref{fig:XXZ}), which locates at $\Delta_c=1$.
It is clear that $G_n$ and $\frac{G_n}{n}$ show a size-independent peak in the vicinity of $\Delta=1$, which indicates large amount of quantum correlation in the QPT.
In a very recent paper studying the multi-partite quantum entanglement for continuous $n$-site subchains of the infinite XXZ model, it has been reported that the measure of multi-partite entanglement
shows a size-independent minima at $\Delta_c=1$, which indicates low hierarchy of multi-partite entanglement in the QPT.
These two completely opposite results suggest that
the QPT in the XXZ model is driven by some non-trivial type of quantum correlation, i.e., it is highly correlated meanwhile low-hierarchically entangled.

\begin{acknowledgments}
The research was supported by the National Natural Science Foundation
of China (Nos. 11204223, 11404250 and 61404095). This work was also
supported by the Talent Scientific Research Foundation of Wuhan Polytechnic
University (Nos. 2011RZ15, 2012RZ09, 2014RZ18).
\end{acknowledgments}


\begin{thebibliography}{References}

\bibitem{BOOK_QPT} S. Sachdev, Quantum Phase Transitions, Cambridge University Press, 1999.

\bibitem{QE_QPT} A.Osterloh, L. Amico, G. Falci, R. Fazio, Nature 416, 608 (2002).
\bibitem{Bell_inequalitiesQPTs_XXZ_model} L. Justino, T. R. de Oliveira, Phys. Rev. A 85, 052128 (2012).
\bibitem{Bell_QPT_models} B. Cakmak, G. Karpat, Z. Gedik, Phys. Lett. A 376, 2982 (2012).
\bibitem{multi_entanglement} O. Ghne, G. Tth, and H. J. Briegel, New J. Phys. 7, 229 (2005).
\bibitem{Camp_GQD_chain} S. Campbell, L. Mazzola, G. De Chiara, T. J. G. Apollaro, F. Plastina, T. Busch, and M. Paternostro, New J. Phys. 15, 043033 (2013).
\bibitem{QE1}J. Eisert, M. Cramer, and M. B. Plenio, Reviews of Modern Physics 82, 277 (2010).





\bibitem{XY3} J. I. Latorre, E. Rico, and G. Vidal, Quantum Inf. Comput. 4, 48 (2004).


\bibitem{MBI1} D. Collins, N. Gisin, S. Popescu, D. Roberts, and V. Scarani, Phys. Rev. Lett. 88, 170405 (2002).
\bibitem{MBI2} J.-D. Bancal, N. Brunner, N. Gisin, and Y.-C. Liang, Phys. Rev. Lett. 106, 020405 (2011).
\bibitem{MBI3} J.-D. Bancal, C. Branciard, N. Gisin, and S. Pironio, Phys. Rev. Lett. 103, 090503 (2009).
\bibitem{MBImy} Zhao-Yu Sun, Yu-Yin Wu, Jian Xu, Hai-Lin Huang, and Bi-Fu Zhan, Bo Wang, Cheng-Bo Duan, Phys. Rev. A 89, 022101 (2014).
\bibitem{tobepub} To be published in Phys. Rev. A.



\bibitem{QD1} H. Ollivier and W. H. Zurek, Phys. Rev. Lett. 88, 017901 (2001).
\bibitem{QD2} T.Werlang, S. Souza, F. F. Fanchini, and C. J. Villas Boas, Phys. Rev. A 80, 024103 (2009).
\bibitem{QD3} B. Wang, Z. Y. Xu, Z. Q. Chen, and M. Feng, Phys. Rev. A 81, 014101 (2010).


\bibitem{QD_chain1} M. S. Sarandy, Phys. Rev. A 80, 022108 (2009).
\bibitem{QD_chain2} R. Dillenschneider, Phys. Rev. B 78, 224413 (2008).
\bibitem{myQDChain} Zhao-Yu Sun, Liang Li,1 Kai-Lun Yao, Gui-Huan Du, Ji-Wei Liu, Bo Luo, Neng Li, and Hai-Na Li, Phys. Rev. A 82, 032310 (2010).
\bibitem{my2D_QDchain} Z.-Y. Sun, L. Li, N. Li, K.-L. Yao, J. Liu, B. Luo, G.-H. Du and H.-N. Li, EPL, 95  30008 (2011).



\bibitem{GQD1} C. C. Rulli and M. S. Sarandy. Phys. Rev. A 84, 042109 (2011).
\bibitem{GQD2} M. Okrasa and Z. Walczak. Europhys. Lett. 96 60003 (2011).
\bibitem{GQD3} D. P. Chi, J. S. Kim, and K. Lee, Phys. Rev. A 87, 062339 (2013).

\bibitem{GQD_mono1} H. C. Braga, C. C. Rulli, Thiago R. de Oliveira, and M. S. Sarandy, Phys. Rev. A 86, 062106 (2012).
\bibitem{GQD_review} Jian-Song Zhang and Ai-Xi Chen, Quant. Phys. Lett. 1, 69  (2012).


\bibitem{GQD_chain} Si-Yuan Liua, Yu-Ran Zhangb, Li-Ming Zhaob, Wen-Li Yanga, Heng Fanb, Annals of Physics, 348, 256 (2014).


\bibitem{MPS} M. M. Wolf, G. Ortiz, F. Verstraete, and J. I. Cirac, Phys. Rev. Lett. 97, 110403 (2006).


\bibitem{iTEBD1} G. Vidal, Phys. Rev. Lett. 91, 147902 (2003).
\bibitem{iTEBD2} B. Pirvu, V. Murg, J. I. Cirac and F. Verstraete, New J. Phys. 12, 025012 (2010).

\bibitem{MPS_faith} F. Verstraete and J. I. Cirac, Phys. Rev. B 73, 094423 (2006).

\bibitem{GQD_analytic} Jianwei Xu, Phys. Lett. A 377,  238 (2013).


\bibitem{DMRG} S. R. White, Phys. Rev. Lett. 69, 2863 (1992).

\bibitem{Cluster} W. Son, L. Amico, R. Fazio, A. Hamma, S. Pascazio, and V. Vedral, EPL 95 50001  (2011).

\bibitem{fac} Salvatore M. Giampaolo, Gerardo Adesso, and Fabrizio Illuminati, Phys. Rev. B 79, 224434 (2009).

\bibitem{CC_SSB}  Olav F. Syljuasen, Phys. Rev. A 68, 060301(R) (2003).

\bibitem{discord_SSB}  Marcelo S. Sarandy, Thiago R. de Oliveira, and Luigi Amico, Int. J. Mod. Phys B 27, 1345030 (2013); Bruno Tomasello et al, Int. J. Mod. Phys. B 26, 1243002 (2012).



\end{thebibliography}
\end{document}